\documentstyle[epsf,preprint,eqsecnum,,aps]{revtex}
\begin{document}
\draft
\title{Magnetically mediated superconductivity:
Crossover from cubic to tetragonal lattice}

\author{P. Monthoux and G.G. Lonzarich}
\address{Cavendish Laboratory, University of Cambridge
\\Madingley Road, Cambridge CB3 0HE, United Kingdom}
\date{\today}
\maketitle

\begin{abstract}

We compare predictions of the mean-field theory of superconductivity for 
nearly-antiferromagnetic and nearly-ferromagnetic metals for cubic and 
tetragonal lattices.  The calculations are based on the parameterisation of 
an effective interaction arising from the exchange of the magnetic 
fluctuations and assume that a single band is relevant for 
superconductivity.  The results show that for comparable model parameters, 
the robustness of magnetic pairing increases gradually as one goes from a 
cubic structure to a more and more anisotropic tetragonal structure either 
on the border of antiferromagnetism or ferromagnetism.

\end{abstract}


\narrowtext

\section{Introduction}

One can expect that the effective interaction between quasiparticles in 
strongly correlated electron systems to be very complex.  The interaction 
will depend obviously on the charge, but also more generally on the spin 
and current carried by the quasiparticles.  On the border of long-range 
magnetic order it is plausible that the dominant interaction channel is of 
magnetic origin and depends on the relative spin orientations of the 
interacting quasiparticles.

It has been shown that this magnetic interaction treated at the mean field 
level can produce anomalous normal state properties and superconducting 
instabilities to anisotropic pairing states.  It correctly predicted the 
symmetry of the Cooper state in the copper oxide 
superconductors\cite{DwavePrediction} and is consistent with 
spin-triplet p-wave pairing in superfluid $^3He$ [for a 
recent review see, e.g., ref.~\cite{Dobbs}].  One also gets the correct 
order of magnitude of the superconducting and superfluid transition 
temperature $T_c$ when the model parameters are inferred from experiments 
in the normal state of the above systems.  There is growing evidence 
that the magnetic interaction model may be relevant to other materials 
on the border of magnetism.

Thus far the magnetic interaction model has been explored in very simple 
cases.  The most extensively investigated example is that of a nearly 
half-filled single-band in a square or cubic lattice.  These studies have 
revealed a number of interesting features that are quite in contrast to 
those expected for conventional phonon mediated pairing.  In the latter 
case, the interaction is local in space, but non-local in time, whereas on 
the border of magnetism, one expects the interaction to be strongly 
non-local in both space and time.  For nearly antiferromagnetic metals the 
magnetic interaction is oscillatory in space and superconductivity depends 
on the ability of the electrons in a Cooper pair state to sample mainly the 
attractive regions of these oscillations.  Because of the strong 
retardation in time, the relative wavefunction of the Cooper pair must be 
constructed from Bloch states with wavevectors close to the Fermi surface. 
 Furthermore, the allowed symmetries of the Cooper pair wavefunction are 
restricted by the crystal structure.  The possibility of constructing a 
Cooper pair state with maximum probability in the attractive regions of the 
magnetic interaction can be severely constrained by these requirements. 
 Therefore, one expects that the robustness of magnetic pairing to be very 
sensitive to details of the electronic and lattice structures.

On the border of ferromagnetism, one is not hampered by the oscillatory 
nature of the magnetic interaction which in the simplest model is 
attractive everywhere in space and time in the spin-triplet channel. At 
first sight this would seem to be the most favourable case for magnetically 
mediated superconductivity.  However, the results of the numerical 
calculations presented in ref.~\cite{ML2} indicate that the highest mean field 
$T_c$ for the cases considered is obtained for d-wave pairing in the nearly 
antiferromagnetic state in a quasi-2D tetragonal lattice.  In this 
particular case, as explained in ref.~\cite{ML2}, it turns out to be 
possible to ideally match the Cooper pair state to the attractive regions 
of the magnetic interaction.

On the border of ferromagnetism, magnetic pairing in the spin-triplet state 
has the disadvantage that only the exchange of magnetic fluctuations 
polarised along the direction of the interacting spins, i.e., longitudinal 
fluctuations, contribute to the quasiparticle interactions.  For a 
spin-rotationally invariant system, both longitudinal and transverse 
fluctuations contribute to pairing only for a spin-singlet state.

Another disadvantage of being on the border of ferromagnetism is that for 
otherwise similar conditions the suppression of $T_c$ due to the self 
interaction arising from the exchange of magnetic fluctuations is stronger 
than in the corresponding case on the border of antiferromagnetism.  This 
disadvantage can be mitigated in systems with strong magnetic anisotropy in 
that the effect of the transverse magnetic fluctuations on the self 
interaction would be suppressed while the strength of the pairing 
interaction arising from the longitudinal magnetic fluctuations need not be 
reduced.  This may apply in systems with strong spin-orbit interactions or 
in the spin-polarised state close to the border of ferromagnetism.

These arguments ~\cite{ML1,ML2} have stimulated a new search for evidence of 
superconductivity on the border of itinerant electron ferromagnetism in 
cases where spin anisotropy is expected to be pronounced, such as $UGe_2$. 
This search has proved fruitful because it led to the first observation of 
the coexistence of superconductivity and itinerant electron ferromagnetism 
in $UGe_2$\cite{Saxena} and shortly thereafter in $ZrZn_2$\cite{Pfleiderer} 
and $URhGe$\cite{Huxley}.

The prediction of the simple model presented in ref.~\cite{ML2} that 
magnetic pairing is more robust in the quasi-2D square lattice than 
in the cubic structure seems to have been borne out by recent experiments.
Namely, one finds an order of magnitude increase in the maximum $T_c$ and 
in the range in pressure where superconductivity is observed on the 
border of metallic antiferromagnetism when the simple cubic lattice of 
$CeIn_3$\cite{Walker} is stretched along one principal axis by the insertion 
of non-magnetic layers to form the tetragonal compounds $CeMIn_5$\cite{Fisk}
($M$ is $Co$, $Rh$ or $Ir$).

These systems, albeit quite anisotropic, would not normally be considered 
to be quasi-two dimensional and it is not clear at first sight that the 
model calculations carried out in ref.~\cite{ML2} are directly relevant. 
The purpose of this paper is to show that for comparable model parameters, 
the robustness of magnetic pairing increases gradually as one goes from 
a cubic structure to a more and more anisotropic structure on the border 
of metallic antiferromagnetism and ferromagnetism.  This behaviour of 
the mean field transition temperature is in stark contrast to that of 
the "one-loop" fluctuations corrections to $T_c$.  The latter corrections 
typically depend logarithmically on the degree of anisotropy and would 
be expected to be negligible for materials such as $CeMIn_5$.

We do not expect some of the results of ref.~\cite{ML2} to be generic 
properties of the magnetic interaction model.  We have already stressed 
that even in simple cases, the robustness of magnetic pairing can be very 
sensitive to certain details of the lattice and electronic structure.  
Even in the single band problem, many such structures have not yet been 
extensively studied theoretically.  Furthermore, we expect that the range 
of possibilities to be greatly expanded in the presence of more than one 
partially filled electronic band.

Most known materials on the border of magnetism crystallise in other than 
simple cubic or tetragonal structure and have more than one band crossing 
the Fermi level.  For these more complex systems, one would not expect the 
model of ref.~\cite{ML2} to be directly relevant.  For example, the 
observation of spin-triplet rather than spin-singlet d-wave pairing in 
some multi-band materials with strongly enhanced antiferromagnetic spin 
fluctuations, such as $UPt_3$ and $Sr_2RuO_4$, may not be inconsistent 
with the idea of magnetic pairing.  A detailed study of magnetic pairing 
in multi-band systems for a range of crystal structures would shed light 
on the possible forms of superconductivity and the conditions most 
favorable for their observation.

The simple model calculations suggest that anisotropic forms of 
superconductivity should be a generic property of systems on the border of 
metallic magnetism.  It may seem surprising therefore that there are still 
so few observations of this phenomenon.  In many cases, the multiplicity of 
bands and, for example, magnetic fluctuations in the non-bipartite lattice 
may weaken magnetic pairing to such an extent that quenched disorder may 
completely suppress superconductivity.  An illustration of this point is 
the dramatic collapse of the spin-triplet superconducting transition 
temperature in $Sr_2RuO_4$ in the presence of $Al$ impurity concentrations 
as low as 0.1\%.

At first sight, the magnetic interaction model is mathematically analogous 
to the conventional electron-phonon problem with the generalised magnetic 
susceptibility playing the role of the phonon propagator.  One would 
therefore expect that a simple analytic expression similar to that proposed 
by McMillan could be used to represent approximately the $T_c$ calculated 
numerically via the Eliashberg equations.  Our attempts in this direction 
have not, however, proved successful\cite{ML1,ML2}.  A recent study 
suggests that there may be a fundamental reason for inapplicability of 
the McMillan-style expression for $T_c$\cite{Abanov}.  On the border of 
long-range magnetic order, the incoherent part of the electron Green 
function, which is ignored in the simplest treatment, plays a major role 
in the formation of the Cooper pair condensate.  The traditional picture 
in which superconductivity arose from pairing of well defined (weakly 
damped) quasiparticles appears inadequate on the border of metallic 
magnetism even in the mean field Eliashberg treatment.

We note that in our model the coupling of the quasiparticles to the 
magnetic fluctuations is a phenomenological constant to be inferred from 
normal state properties that formally includes that part of the vertex 
correction which is local in space and time.  Calculations have shown that 
the neglect of vertex corrections that are non-local in space and time is 
justified at least in some cases of physical interest\cite{Vertex}.  
When the magnetic correlation length becomes sufficiently large, however, 
these neglected non-local vertex corrections (including superconducting 
phase fluctuations) may become important.  Their effect on $T_c$ and 
the normal state properties are as yet incompletely understood.

\section{Model}

We consider quasiparticles in a simple tetragonal lattice described by a 
dispersion relation

\begin{eqnarray}
\epsilon_{\bf p} & = & -2t(\cos(p_x) + \cos(p_y) + \alpha_t\cos(p_z)) 
\nonumber \\ & - & 4t'(\cos(p_x)\cos(p_y) + \alpha_t\cos(p_x)\cos(p_z) 
+ \alpha_t\cos(p_y)\cos(p_z))
\label{eps}
\end{eqnarray}

\noindent with hopping matrix elements $t$ and $t'$. $\alpha_t$ represents 
the electronic structure anisotropy along the z direction.  
$\alpha_t = 0$  corresponds to the quasi-2D limit while 
$\alpha_t = 1$ corresponds to the 3D cubic lattice.  For 
simplicity, we measure all lengths in units of the respective lattice 
spacing.  In order to reduce the number of independent parameters, we take 
$t' = 0.45 t$ and a band filling factor $n = 1.1$ as in our earlier work.

The effective interaction between quasiparticles is assumed to be isotropic 
in spin space and is defined in terms of the coupling constant $g$ and the 
generalised magnetic susceptibility which is assumed to have a simple 
analytical form consistent with the symmetry of the lattice.

\begin{equation}
\chi({\bf q},\omega) = {\chi_0\kappa_0^2\over \kappa^2 + \widehat{q}^2 
- i{\omega\over \eta(\widehat{q})}}
\label{chiML}
\end{equation}

\noindent where $\kappa$ and $\kappa_0$ are the correlation wavevectors or 
inverse correlation lengths in units of the lattice spacing in the basal 
plane, with and without strong magnetic correlations, respectively.  Let

\begin{equation}
\widehat{q}_{\pm}^2 = (4 + 2\alpha_m) \pm 
2(\cos(q_x)+\cos(q_y)+\alpha_m\cos(q_z)) 
\label{qdef}
\end{equation}

\noindent where $\alpha_m$ parameterises the magnetic anisotropy. $\alpha_m = 0$ 
corresponds to quasi-2D magnetic correlations and $\alpha_m = 1$ 
corresponds to 3D magnetic correlations.

In the case of a nearly ferromagnetic metal the parameters $\widehat{q}^2$
and $\eta(\widehat{q})$ in Eq.~(\ref{chiML}) are defined as

\begin{eqnarray}
\widehat{q}^2 & = & \widehat{q}_{-}^2 \\
\eta(\widehat{q}) & = & T_{sf}\widehat{q}_{-}
\label{ferro}
\end{eqnarray}

\noindent where $T_{sf}$ is a characteristic spin fluctuation temperature.  
Note that our definition of $T_{sf}$ may differ from the characteristic 
spin fluctuation temperature scales used by other authors.

In the case of a nearly antiferromagnetic metal, the parameters $\widehat{q}^2$
and $\eta(\widehat{q})$ in Eq.~(\ref{chiML}) are defined as

\begin{eqnarray}
\widehat{q}^2 & = & \widehat{q}_{+}^2 \\
\eta(\widehat{q}) & = & T_{sf}\widehat{q}_{-}
\label{antiferro}
\end{eqnarray}

As in our previous work\cite{ML1,ML2}, the band structure and generalized 
magnetic susceptibility are modeled independently. This choice may be 
inconsistent when all of the contributions to $\chi({\bf q},\omega)$ 
come from the chosen band. However, it allows us, in principle, to deal 
with the case where there are other important contributions to the 
generalized magnetic susceptibility. It has been argued that the latter 
case is of relevance to the ruthenates\cite{Kuwabara}, and most likely
the heavy-fermion systems.

A complete description of the model, the Eliashberg equations for the 
superconducting transition temperature and their method of solution 
can be found in the appendix.

We note that the model is fully defined by the phenomenological parameters 
describing the electronic structure $\epsilon_{\bf p}$, the generalised 
magnetic susceptibility $\chi({\bf q},\omega)$ and the interaction 
vertex $g$.  In principle, these parameters can be estimated from 
experimental studies of the normal state. In particular, the resistivity 
can be used to estimate the dimensionless coupling parameter $g^2\chi_0/t$ 
the value of which is between 10 and 20 for the simplest RPA model for 
the magnetic interaction.

\section{Results}

\subsection{Solution of the Eliashberg Equations for $T_c$}

The dimensionless parameters at our disposal are $g^2\chi_0/t$, 
$T_{sf}/t$, $\kappa_0$ and $\kappa$.  For comparison with the 
results of our earlier work\cite{ML1,ML2}, we take $T_{sf} = {2\over 3} t
$ and $\kappa_0^2 = 12$. In 2D, this $T_{sf}$ corresponds to about 
$1000^\circ K$ for a bandwidth of 1 eV while our choice of of 
$\kappa_0^2 $ is a representative value.

The results of our numerical calculations of the mean field critical 
temperature $T_c$ as a function of the electronic and magnetic anisotropy 
parameters $\alpha_t$ and $\alpha_m$, respectively, are shown in Figs 1 and 2 
for representative values of the parameters $\kappa^2$ and $g^2\chi_0/t$.  
Figures 1a-c illustrate the results for a nearly anti-ferromagnetic metal 
and figures 2a-c for a nearly ferromagnetic metal.  Note that our previous 
calculations correspond to the quasi-2D case $\alpha_t = \alpha_m = 0$ 
and to 3D case $\alpha_t = \alpha_m = 1$.

A glance at figure 1 reveals a clear pattern in the variation of $T_c$ 
with both anisotropy parameters $\alpha_t$ and $\alpha_m$.  We notice 
that $T_c$ increases gradually and monotonically as the system becomes 
more and more anisotropic in either the electronic structure or in the 
magnetic interaction. In going from 3D to quasi-2D, $T_c/T_{sf}$ is found 
to increase by up to an order of magnitude for otherwise fixed parameters 
of the model. The increase becomes least pronounced as for small 
$\kappa^2$ and large $g^2\chi_0/t$.

The behavior in the nearly ferromagnetic case, figure 2, though broadly 
similar to that of the nearly antiferromagnetic metal, shows some 
interesting differences.  In some cases, the minimum $T_c$ occurs for 3D 
electronic structure, but quasi-2D magnetic interaction.  Also, in all 
cases considered the maximum $T_c$ is obtained for a quasi-2D electronic 
structure and strongly anisotropic, but not 2D magnetic interactions.

\subsection{Mass renormalisation and interaction parameter}

In order to make a comparison with the corresponding electron-phonon 
problem it is instructive to define a mass renormalization parameter 
$\lambda_Z$ and interaction parameter $\lambda_\Delta$. We define

\begin{eqnarray}
\lambda_Z & = & { \int_{-\infty}^{+\infty} {d\omega\over \pi}
<{1\over \omega}ImV_Z({\bf p}-{\bf p'},\omega)>_{FS({\bf p},{\bf p'})} 
\over <1>_{FS({\bf p})} } \label{lambda1} \\
\lambda_\Delta & = & -{ \int_{-\infty}^{+\infty} {d\omega\over \pi}
<{1\over \omega}ImV_\Delta({\bf p}-{\bf p'},\omega)
\eta({\bf p})\eta({\bf p'})>_{FS({\bf p},{\bf p'})} 
\over <\eta^2({\bf p})>_{FS({\bf p})} } \label{lambda2}
\end{eqnarray}

\noindent where

\begin{equation}
V_Z({\bf q},\omega) = g^2\chi({\bf q},\omega)
\label{Vz}
\end{equation}

\noindent and

\begin{eqnarray}
V_p({\bf q},\omega) & = & -{g^2\over 3}\chi({\bf q},\omega) \\
\eta({\bf p}) & = & \sin(p_x)
\label{Vp}
\end{eqnarray}

\noindent for p-wave spin triplet pairing $(\Delta \equiv p)$ while

\begin{eqnarray}
V_d({\bf q},\omega) & = & g^2\chi({\bf q},\omega) \\
\eta({\bf p}) & = & \cos(p_x) - \cos(p_y)
\label{Vd}
\end{eqnarray}

\noindent in the case of d-wave spin-singlet pairing $(\Delta \equiv d)$. 
The Fermi surface averages are given by

\begin{eqnarray}
<\cdots>_{FS({\bf p})}  & = & \int {d^dp\over (2\pi)^d} \cdots 
\delta(\epsilon_{\bf p} - \mu) \label{FSaverage1} \\
<\cdots>_{FS({\bf p},{\bf p'})}  & = & \int {d^dp\over (2\pi)^d} 
{d^dp'\over (2\pi)^d}\cdots \delta(\epsilon_{\bf p} - \mu) 
\delta(\epsilon_{\bf p'} - \mu)  \label{FSaverage2}
\end{eqnarray}

\noindent In practice, we compute the Fermi surface average with a 
discrete set of momenta on a cubic or tretragonal lattice and we 
replace the delta function by a finite temperature expression

\begin{eqnarray}
\int {d^dp\over (2\pi)^d} & \longrightarrow &{1\over N}\sum_{\bf p} \\
\delta(\epsilon_{\bf p} - \mu) & \longrightarrow & {1\over T} f_{\bf p}(1-f_
{\bf p})
\label{FSaverageNum}
\end{eqnarray}

\noindent where $f_{\bf p}$ is the Fermi function. Note that 
${1\over T} f_{\bf p}(1-f_{\bf p}) \rightarrow 
\delta(\epsilon_{\bf p} - \mu)$ as $T \rightarrow 0$. We have used 
$T = 0.1t$ and $N = 128^d$ in all of our calculations. The finite 
temperature effectively means that van Hove singularities will be 
smeared out.

Note that the Fermi surface average that appears in $\lambda_Z$, 
Eq.~(\ref{lambda1}) plays a role similar to that of 
$\alpha^2F(\omega)/\omega$ in the case of phonon mediated 
superconductivity. From the definitions of the parameters 
$\lambda_{Z,\Delta}$ Eqs.~(\ref{lambda1}), ~(\ref{lambda2}) and our 
model for $\chi({\bf q},\omega)$ Eq.~(\ref{chiML}), we see 
that $\lambda_{Z,\Delta}$ are directly proportional to the dimensionless 
factor $g^2\chi_0\kappa_0^2/t$. Thus we will consider the quantities

\begin{equation}
\lambda^*_{Z,\Delta} \equiv \lambda_{Z,\Delta}/(g^2\chi_0\kappa_0^2/t)
\label{lambdaBar}
\end{equation}

\noindent which are functions only of $n$, $t'/t$ and $\kappa^2$.

In figures 3 and 4 we show $\lambda_Z^*$, $\lambda_\Delta^*$ and the 
ratio $\lambda_\Delta/\lambda_Z$ for a representative 
value of $\kappa^2$ in the case of a nearly antiferromagnetic metal 
and nearly ferromagnetic metal, respectively.

The trends in both cases are the same.  $\lambda_Z^*$ and 
$\lambda_\Delta^*$ are seen to increase gradually and monotonically 
in going from 3D to quasi-2D.  However, $\lambda_\Delta^*$ grows faster 
than $\lambda_Z^*$ so the ratio $\lambda_\Delta/\lambda_Z$ also 
increases in going from 3D to quasi-2D.  This qualitative trend 
in the ratio is consistent with the behavior of $T_c$ obtained from the 
numerical solution of the Eliashberg equations.  In the ferromagnetic 
case, however, it fails to reproduce the fact that the minimum $T_c$
is not necessary for a fully 3D system and that the maximum $T_c$ is 
obtained for strongly anisotropic yet not quasi-2D systems.

\section{Discussion}

The results of the calculations for both the nearly ferromagnetic and 
nearly antiferromagnetic metals show that the robustness of magnetic 
pairing increases gradually as one goes from a cubic to a more and more 
anisotropic structure with parameters other than $\alpha_m$ and $\alpha_t$ 
left unchanged. These results are consistent with our previous findings~\cite{ML2}
and with the calculations for $\alpha_m = \alpha_t=0$ and $\alpha_m = \alpha_t=1$
presented in ref.~\cite{Arita}. In an earlier study, Nakamura et al.
~\cite{Nakamura} found that $T_c$ could increase by up to a factor of three
in going from 3D to 2D for their choice of model parameters. The effect of 
anisotropy on $T_c$ for nearly ferromagnetic and nearly antiferromagnetic 
metals is qualitatively similar. This phenomenon 
arises from the increase with growing anisotropy of the 
density of states of both the quasiparticles and of the magnetic 
fluctuations that mediate the quasiparticle interaction.  This effect could be 
further enhanced in the case of a nearly antiferromagnetic metal by the 
change in the pattern of the oscillations of the magnetic interaction.

It can be seen from figure 5 that the strength of the interaction in the 
repulsive sites outside of the nodal plane of the $d_{x^2-y^2}$ state gets 
reduced while crucially the attraction in the basal plane gets enhanced 
as one goes from the cubic to a more and more anisotropic tetragonal 
lattice.  This enhancement is the consequence of the increase of the phase 
space of soft magnetic fluctuations as one goes from a cubic to a 
quasi-two dimensional structure.  Since our model potential varies 
smoothly with the tetragonal distortion, parameterised by $\alpha_m$ 
in figures 1 and 2, it is clear that these effects occur gradually 
with increasing separation between the basal planes.

The calculations assume that the maximum magnetic response for a nearly 
antiferromagnetic metal occurs at the commensurate wavevector defined 
by $Q_x = Q_y = \pi/a$ and $Q_z = \pi/c$, where $a$ and $c$ are the lattice
constants in the basal plane and along the tetragonal axis respectively,
reintroduced here for clarity. The oscillations in the 
magnetic interaction potential along the tetragonal axis obviously 
depend on the value of $Q_z$. However, the enhancement of the attraction 
in the basal plane and the reduction of the interaction elsewhere as one 
goes from a cubic to a more and more anisotropic lattice do not depend 
on the particular value of $Q_z$. Therefore, we expect the qualitative 
conclusions of this paper to be independent of $Q_z$.

The robustness of the pairing is further enhanced by the gradual change in 
the electronic band from a 3D to a quasi-2D form (see Eq.~(\ref{eps})).  
The reduced hopping along the distortion axis, parameterised by $\alpha_t$
in figures 1 and 2, implies a reduced electronic bandwidth and hence 
increased density of electronic states.  Our calculations show that this 
too leads to a gradual increase in $T_c$ with increasing distortion of 
the lattice.

In a nearly ferromagnetic metal, one again benefits from the reduction of 
the electronic band width and the increase of the interaction in the basal 
plane as one goes from a cubic to a tetragonal lattice (see figure 6). 
However, the suppression of the interaction between the basal planes has a 
less dramatic effect on the border of ferromagnetism than 
antiferromagnetism because in the latter case one suppresses key repulsive 
regions of the interaction (figure 5).

These simple arguments explain how the pairing effects of the interaction 
are strengthened by a tetragonal distortion in our model.  However, the 
same effects also contribute to an enhanced self-interaction which acts to 
suppress $T_c$.  The relative importance of the pair forming and pair 
breaking effects of the magnetic interaction cannot be inferred by the 
above physical picture alone.  The numerical calculations show that for 
most cases considered here the pair forming effects dominate.  The balance 
is particularly delicate on the border of ferromagnetism where the 
suppression of $T_c$ brought about by the self-interaction is pronounced.  
A physical interpretation of this suppression of $T_c$ is given in 
ref.~\cite{Abanov}.  The same interpretation may explain, for example, 
why the maximum of $T_c/T_{sf}$ in the nearly ferromagnetic case is for 
a strongly anisotropic yet not quasi-2D pairing potential (figure 2).

A most striking manifestation of the interplay between the pair-forming and 
pair-breaking tendency of the magnetic interaction is the breakdown of the 
McMillan-style expression for $T_c$ in terms of the parameters 
$\lambda_\Delta$ and $\lambda_Z$ (see Eqs.~(\ref{lambda1},\ref{lambda2}).  
This was noted in ref.~\cite{ML2} and has been interpreted in 
ref.~\cite{Abanov} in terms of the important role played by the 
incoherent part of the Green function which is ignored in the simplest 
treatments, but is included in the present and earlier work\cite{ML1,ML2}
where the full momentum and frequency dependence of the self 
energy is taken into account.

\section{Outlook}

The calculations show that the lattice anisotropy may increase the 
robustness of magnetic pairing in the mean-field approximation.
Superconducting phase fluctuations which are not included in this
approximation may be expected to suppress $T_c$ in the 2D limit.
Therefore, in practice, one would think that the most favorable
case for magnetic pairing is that of strong but not extreme 
anisotropy.

As noted in the introduction and in the 
previous two sections, the robustness of magnetic pairing can be very 
sensitive to certain details of the magnetic interaction and electronic 
structure.  Therefore, one should exercise caution in making quantitative 
comparisons between the results of our calculations and experiment.  For 
instance, one would expect all of the parameters of the model (not solely 
$\alpha_m$ and $\alpha_t$) to change simultaneously with increasing 
lattice anisotropy. The changes brought about in going from a cubic to 
a tetragonal lattice may even be much more complex than considered here.  
In particular, the number of partially filled bands may itself change.  
As also mentioned in the introduction, this could have in some cases 
even more dramatic consequences on superconductivity than the effects 
taken into account in our simple one-band model.

The theoretical framework developed for systems on the border of 
magnetism can be translated to describe systems on the border of other
types of instabilities, such as charge density wave or ferroelectric 
instabilities.  The above given phase space argument to explain the 
increased robustness of magnetic pairing with increasing lattice 
anisotropy should carry over in part to these other pairing mechanisms,
at least at the one-loop mean-field level (see, e.g., 
ref.~\cite{Ashcroft}).

While some understanding of the properties of the magnetic interaction 
model has been gained over the last few years (e.g., the conditions for 
robust pairing of electrons), there are many cases where the predictions of 
the model have not been worked out.  Of particular importance is the role 
of the multiplicity of partially filled bands which may be expected to be 
the key to understanding exotic superconductivity observed in nearly 
magnetic materials such as $UPt_3$ and $Sr_2RuO_4$.

\section{Acknowledgments}

We would like to thank A.V. Chubukov, P. Coleman, S.R. Julian, 
P.B. Littlewood, A.J. Millis, A.P. Mackenzie, D. Pines, D.J. Scalapino 
and M. Sigrist for discussions on this and related topics. We 
acknowledge the support of the EPSRC, the Newton Trust and the Royal Society.

\section{Appendix}

We consider quasiparticles on a cubic 
or tetragonal lattice. We assume that the dominant scattering 
mechanism is of magnetic origin and postulate the following low-energy 
effective action for the quasiparticles:

\begin{eqnarray}
S_{eff} & = & \sum_{{\bf p},\alpha}\int_0^\beta d\tau 
\psi^\dagger_{{\bf p},\alpha}(\tau)\Big(\partial_\tau + 
\epsilon_{\bf p} - \mu\Big) \psi_{{\bf p},\alpha}(\tau) \nonumber\\
& & - {g^2\over 6 N}\sum_{\bf q}\int_0^\beta 
d\tau \int_0^\beta d\tau' \chi({\bf q},\tau-\tau')
{\bf s}({\bf q},\tau)\cdot {\bf s}(-{\bf q},\tau')
\label{Seff}
\end{eqnarray}

\noindent where $N$ is the number of allowed wavevectors in the Brillouin
Zone and the spin density ${\bf s}({\bf q},\tau)$ is given by

\begin{equation}
{\bf s}({\bf q},\tau) \equiv \sum_{{\bf p},\alpha,\gamma} 
\psi^\dagger_{{\bf p} + {\bf q},\alpha}(\tau){\bf \sigma}_{\alpha,\gamma}
\psi_{{\bf p},\gamma}(\tau)
\label{Spin}
\end{equation}

\noindent where $\bf \sigma$ denotes the three Pauli matrices. The 
quasiparticle dispersion relation $\epsilon_{\bf p}$ is defined in 
Eq.~(\ref{eps}), $\mu$ denotes the chemical potential, $\beta$ the 
inverse temperature, $g^2$ the coupling constant and 
$\psi^\dagger_{{\bf p},\sigma}$ and $\psi_{{\bf p},\sigma}$ are
Grassmann variables. In the following we shall measure temperatures, 
frequencies and energies in the same units. 

The retarded generalized magnetic susceptibility $\chi({\bf q},\omega)$ 
that defines the effective interaction, Eq.~(\ref{Seff}), is defined
in Eq.~(\ref{chiML}).

The spin-fluctuation propagator on the imaginary axis, 
$\chi({\bf q},i\nu_n)$ is related to the imaginary part of the 
response function $Im\chi({\bf q},\omega)$, Eq.~(\ref{chiML}), 
via the spectral representation

\begin{equation}
\chi({\bf q},i\nu_n) = -\int_{-\infty}^{+\infty}{d\omega\over \pi}
{Im\chi({\bf q},\omega)\over i\nu_n - \omega}
\label{chi_mats}
\end{equation}

\noindent To get $\chi({\bf q},i\nu_n)$ to decay as $1/\nu_n^2$ as
$\nu_n \rightarrow \infty$, as it should, we introduce a cutoff 
$\omega_0$ and take $Im\chi({\bf q},\omega) = 0$ for $\omega 
\geq \omega_0$. A natural choice for the cutoff is $\omega_0 
= \eta(\widehat{q})\kappa_0^2$. We have checked that our results for
the critical temperature are not sensitive to the particular choice 
of $\omega_0$ used.

The Eliashberg equations for the critical 
temperature $T_c$ in the Matsubara representation reduce, for the
effective action Eq.~(\ref{Seff}), to

\begin{equation}
\Sigma({\bf p},i\omega_n) = g^2{T\over N}\sum_{\Omega_n}
\sum_{\bf k}\chi({\bf p}-{\bf k},i\omega_n-i\Omega_n)
G({\bf k},i\Omega_n)
\label{Sigma}
\end{equation}

\begin{equation}
G({\bf p},i\omega_n) = {1\over i\omega_n - (\epsilon_{\bf p}-\mu) 
- \Sigma({\bf p} ,i\omega_n)}
\label{Green}
\end{equation}

\begin{eqnarray}
\Lambda(T)\Phi({\bf p},i\omega_n) & = & 
\Bigg[{{g^2\over 3} \atop -g^2}\Bigg]
{T\over N}\sum_{\Omega_n}\sum_{\bf k}
\chi({\bf p} - {\bf k},i\omega_n - i\Omega_n)
|G({\bf k},i\Omega_n)|^2 \Phi({\bf k},i\Omega_n) \nonumber \\
\Lambda(T) & = & 1 \longrightarrow T = T_c
\label{Gap}
\end{eqnarray}

\noindent where $\Sigma({\bf p},i\omega_n)$ is the quasiparticle 
self-energy, $G({\bf p},i\omega_n)$ the one-particle Green's 
function and $\Phi({\bf p},i\omega_n)$ the anomalous self-energy.
The chemical potential is adjusted to give an electron 
density of $n = 1.1$, and $N$ is the total
number of allowed wavevectors in the Brillouin Zone. In Eq.~(\ref{Gap}),
the prefactor $g^2/3$ is for triplet pairing while the
prefactor $-g^2$ is appropriate for singlet pairing. Only the
longitudinal spin-fluctuation mode contributes to the pairing 
amplitude in the triplet channel. Both transverse and longitudinal 
spin-fluctuation modes contribute to the pairing amplitude in the 
singlet channel. All three modes contribute to the quasiparticle 
self-energy.

The momentum convolutions in Eqs.~(\ref{Sigma},\ref{Gap}) are 
carried out with a Fast Fourier Transform algorithm on a 
$48 \times 48 \times 48$ lattice. The frequency sums in both the  
self-energy and linearized gap equations are treated with the 
renormalization group technique of Pao and Bickers\cite{PaoBickers}. 
We have kept between 8 and 16 Matsubara frequencies at each stage of 
the renormalization procedure, starting with an initial temperature 
$T_0 = 0.6t$ and cutoff $\Omega_c \approx 30t$.The renormalization 
group acceleration technique restricts one to a discrete set of 
temperatures $T_0 > T_1 > T_2 \dots$. The critical temperature at 
which $\Lambda(T) = 1$ in Eq.~(\ref{Gap}) is determined by 
linear interpolation.

\begin{figure}
\centerline{\epsfysize=6.00in
\epsfbox{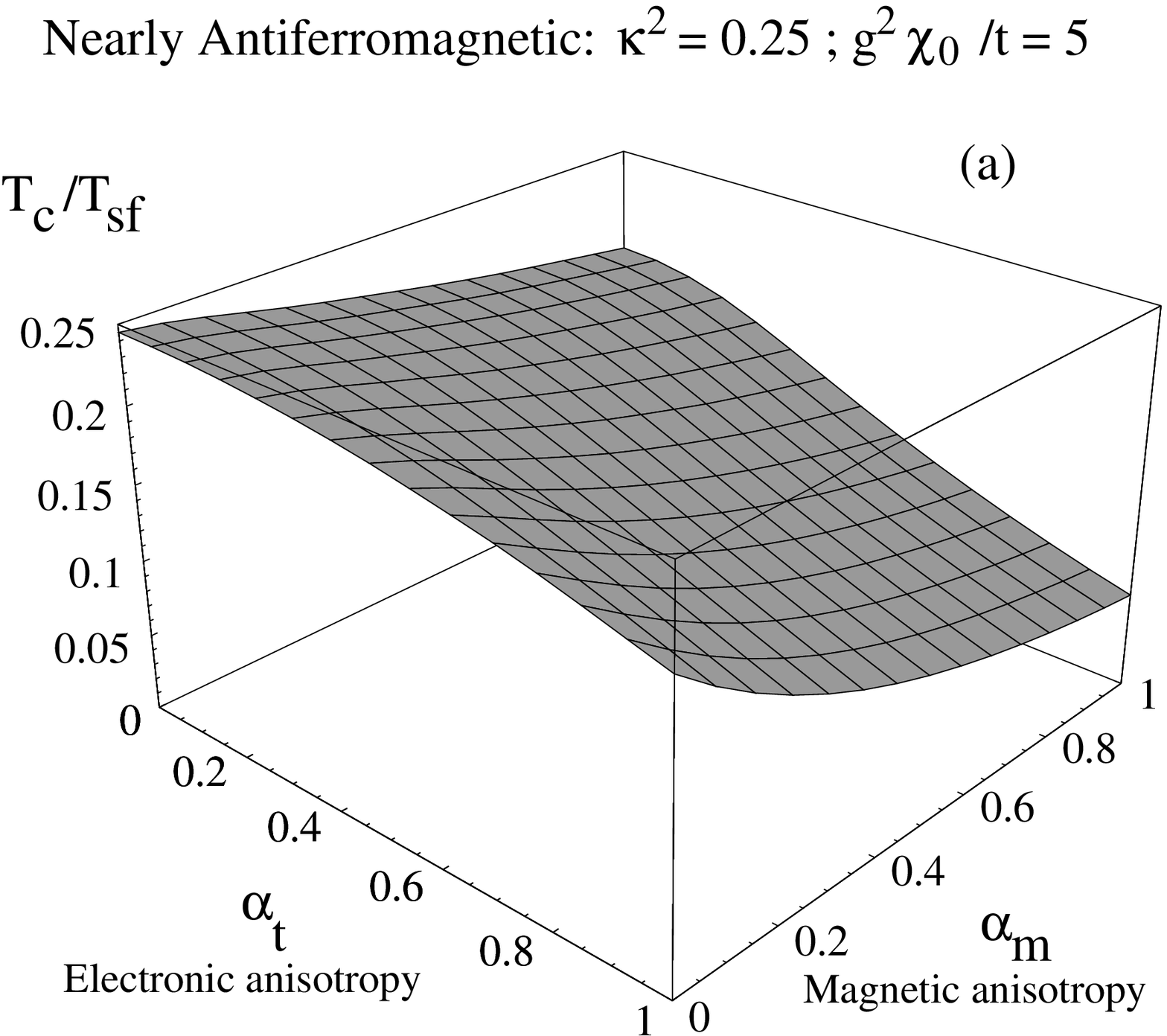}}
\centerline{\epsfysize=6.00in
\epsfbox{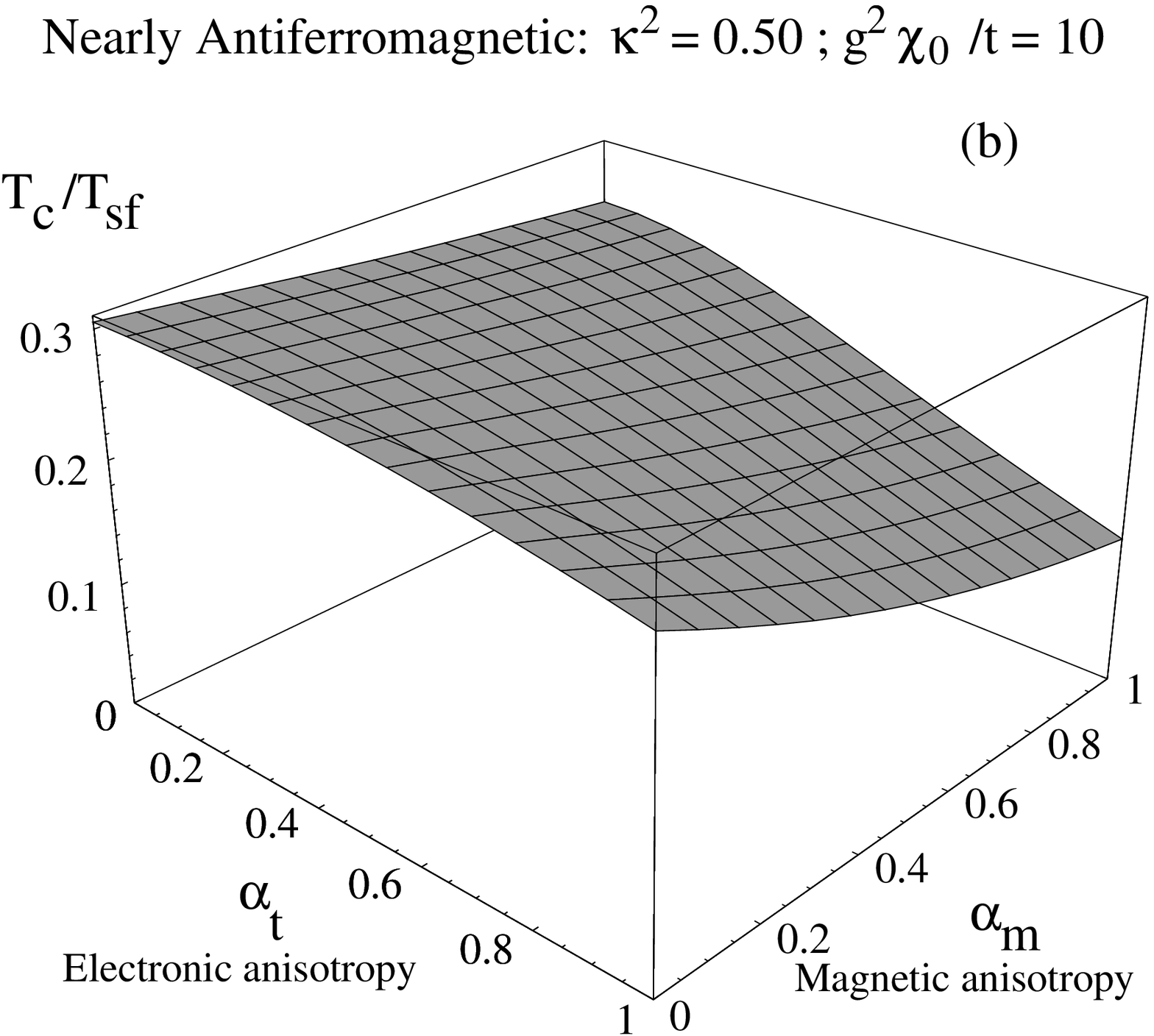}}
\centerline{\epsfysize=6.00in
\epsfbox{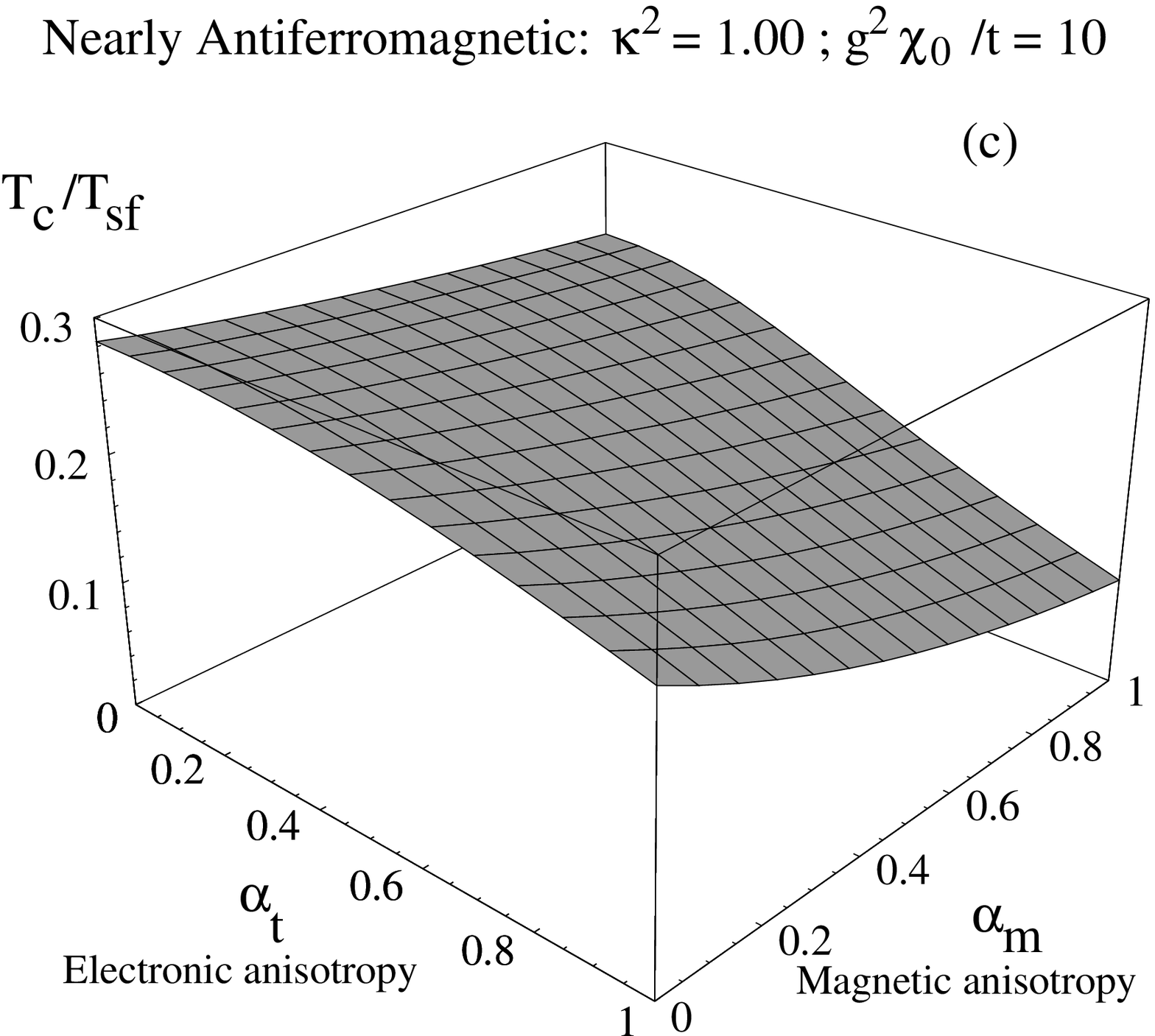}}
\vskip 1.0cm
\caption{Eliashberg $T_c/T_{sf}$ for nearly antiferromagnetic systems 
as a function of the electronic anisotropy parameter $\alpha_t$ 
and the magnetic anisotropy parameter $\alpha_m$ for representative 
values of the correlation wavevector $\kappa^2$ and coupling constant 
$g^2\chi_0/t$. (a) $\kappa^2 = 0.25$, $g^2\chi_0/t = 5$. 
(b) $\kappa^2 = 0.50$, $g^2\chi_0/t = 10$. 
(c) $\kappa^2 = 1.00$, $g^2\chi_0/t = 10$. 
$\alpha_t = \alpha_m = 0$ corresponds to the 2D limit
while $\alpha_t = \alpha_m = 1$ corresponds to an isotropic 3D system.}
\label{fig1}
\end{figure}

\begin{figure}
\centerline{\epsfysize=6.00in
\epsfbox{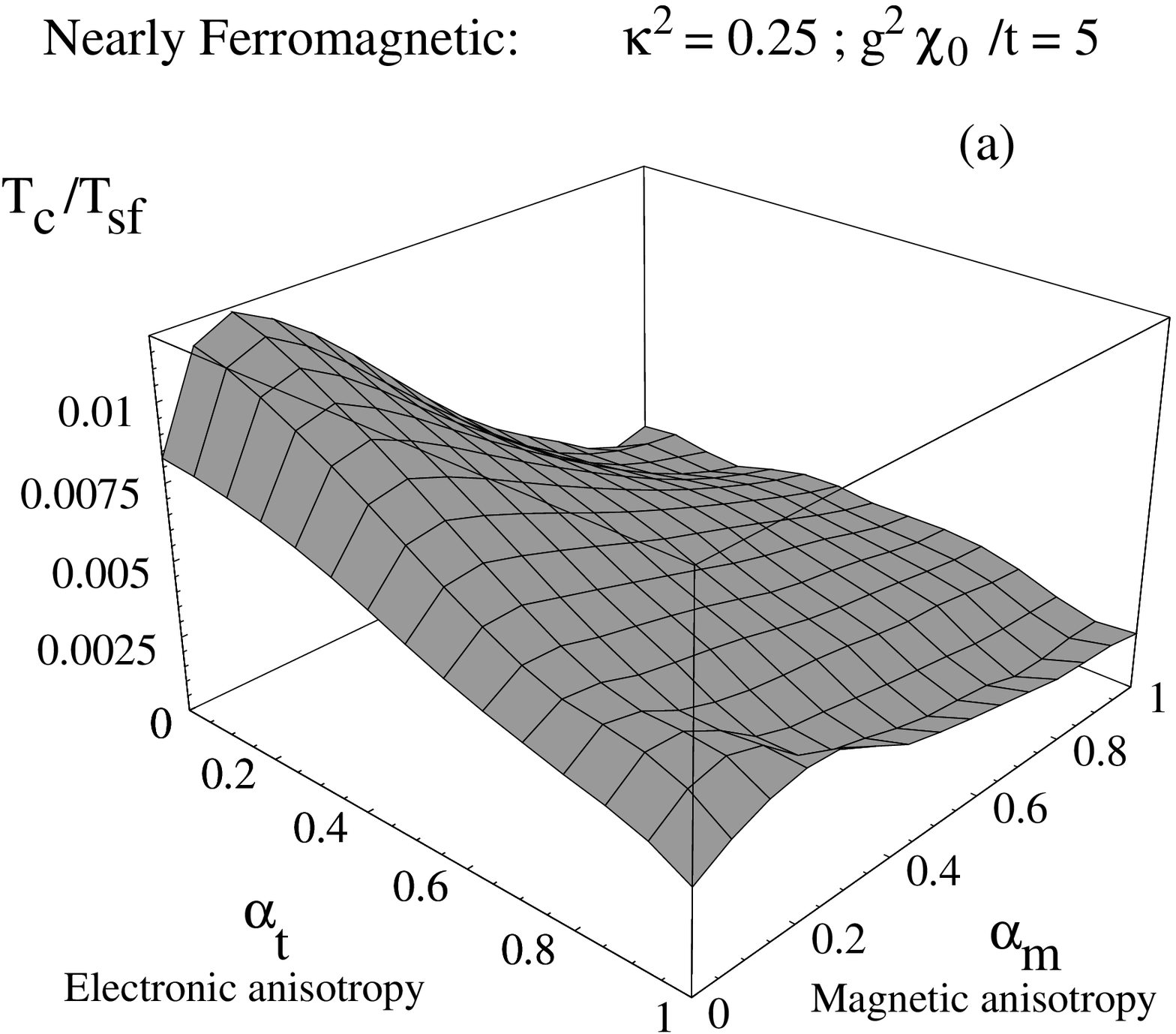}}
\centerline{\epsfysize=6.00in
\epsfbox{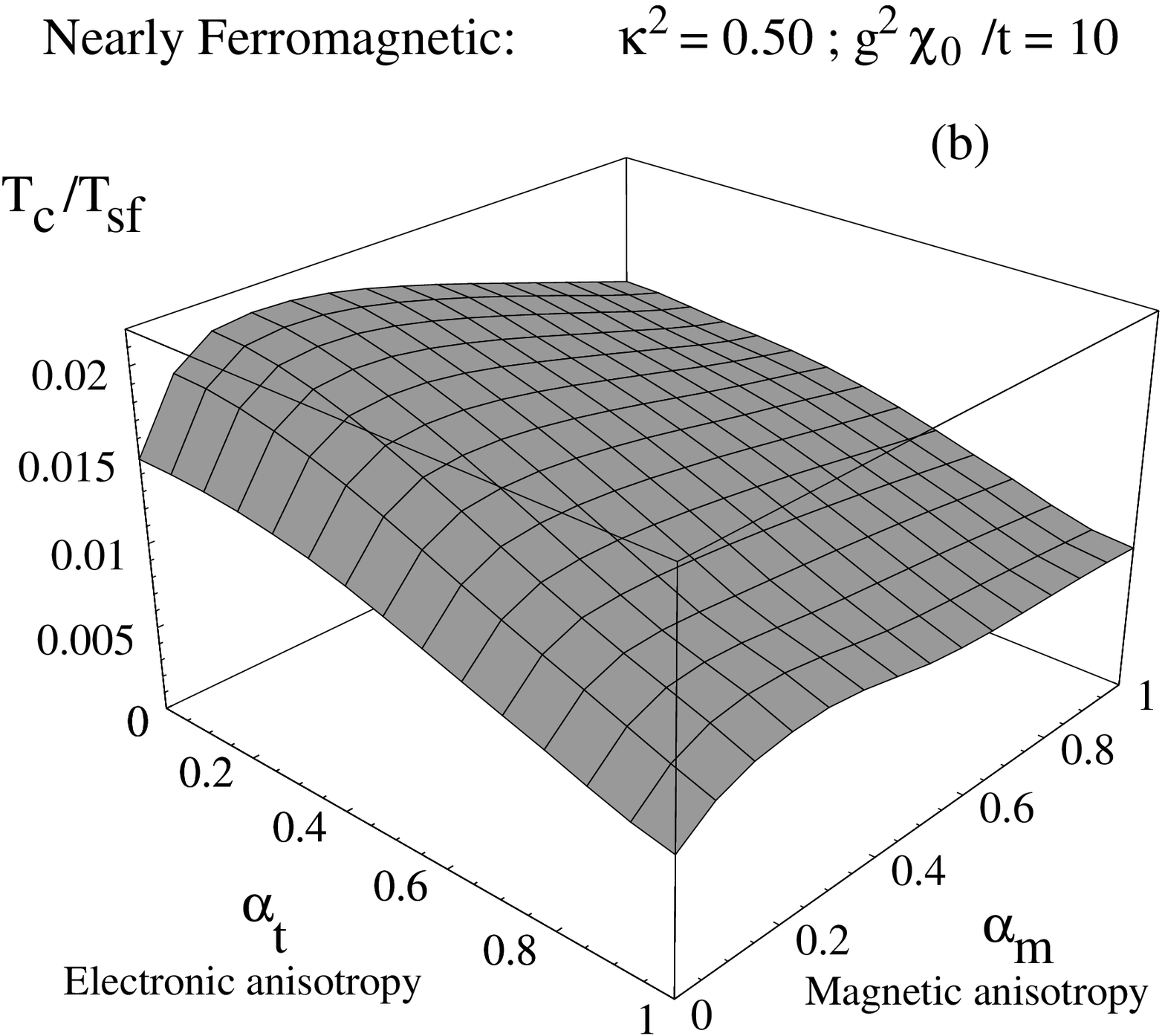}}
\centerline{\epsfysize=6.00in
\epsfbox{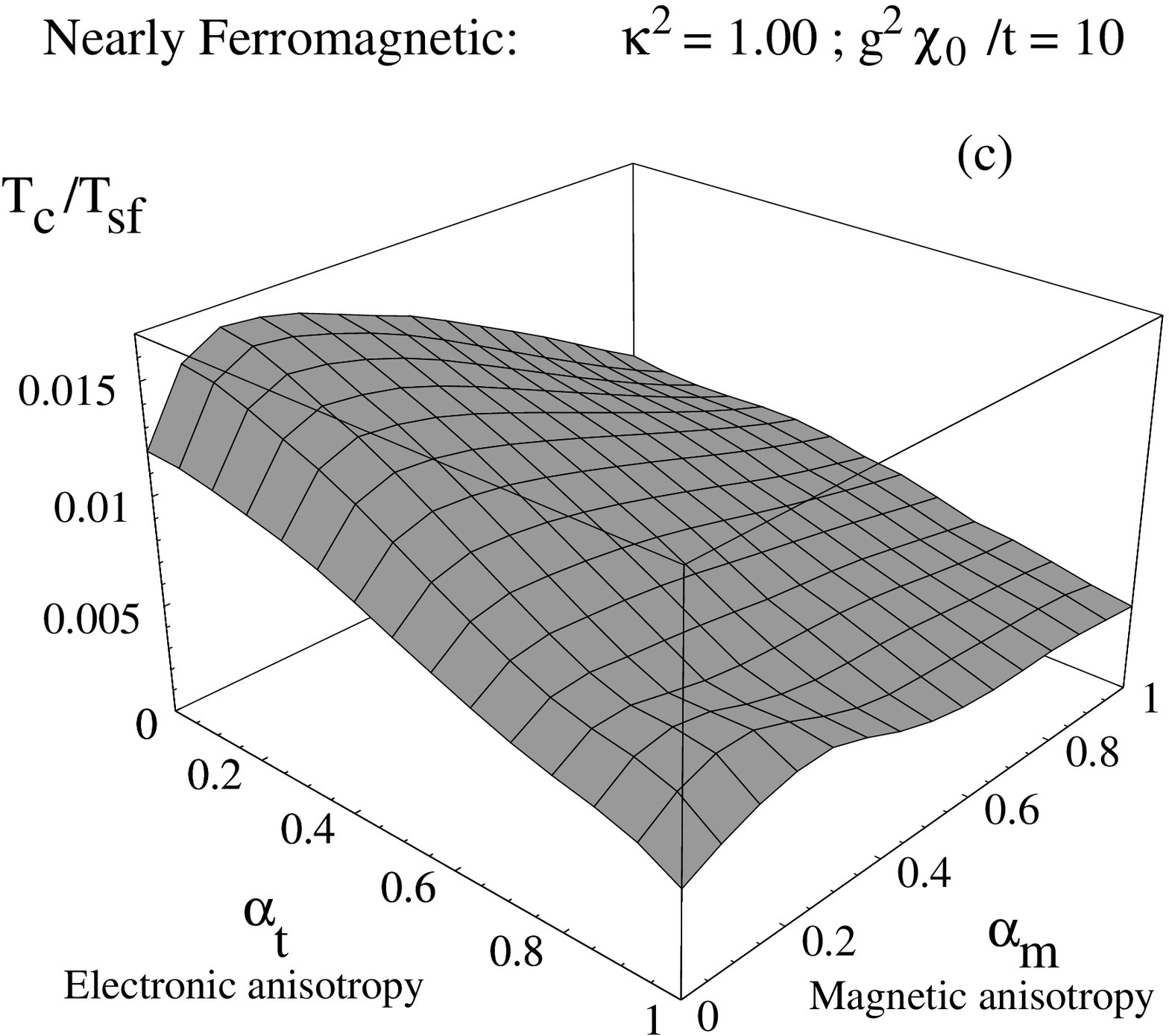}}
\vskip 1.0cm
\caption{Eliashberg $T_c/T_{sf}$ for nearly ferromagnetic systems 
as a function of the electronic anisotropy parameter $\alpha_t$ 
and the magnetic anisotropy parameter $\alpha_m$ for representative 
values of the correlation wavevector $\kappa^2$ and coupling constant 
$g^2\chi_0/t$. (a) $\kappa^2 = 0.25$, $g^2\chi_0/t = 5$. 
(b) $\kappa^2 = 0.50$, $g^2\chi_0/t = 10$. 
(c) $\kappa^2 = 1.00$, $g^2\chi_0/t = 10$. 
$\alpha_t = \alpha_m = 0$ corresponds to the 2D limit
while $\alpha_t = \alpha_m = 1$ corresponds to an isotropic 3D system.}
\label{fig2}
\end{figure}

\begin{figure}
\centerline{\epsfysize=6.00in
\epsfbox{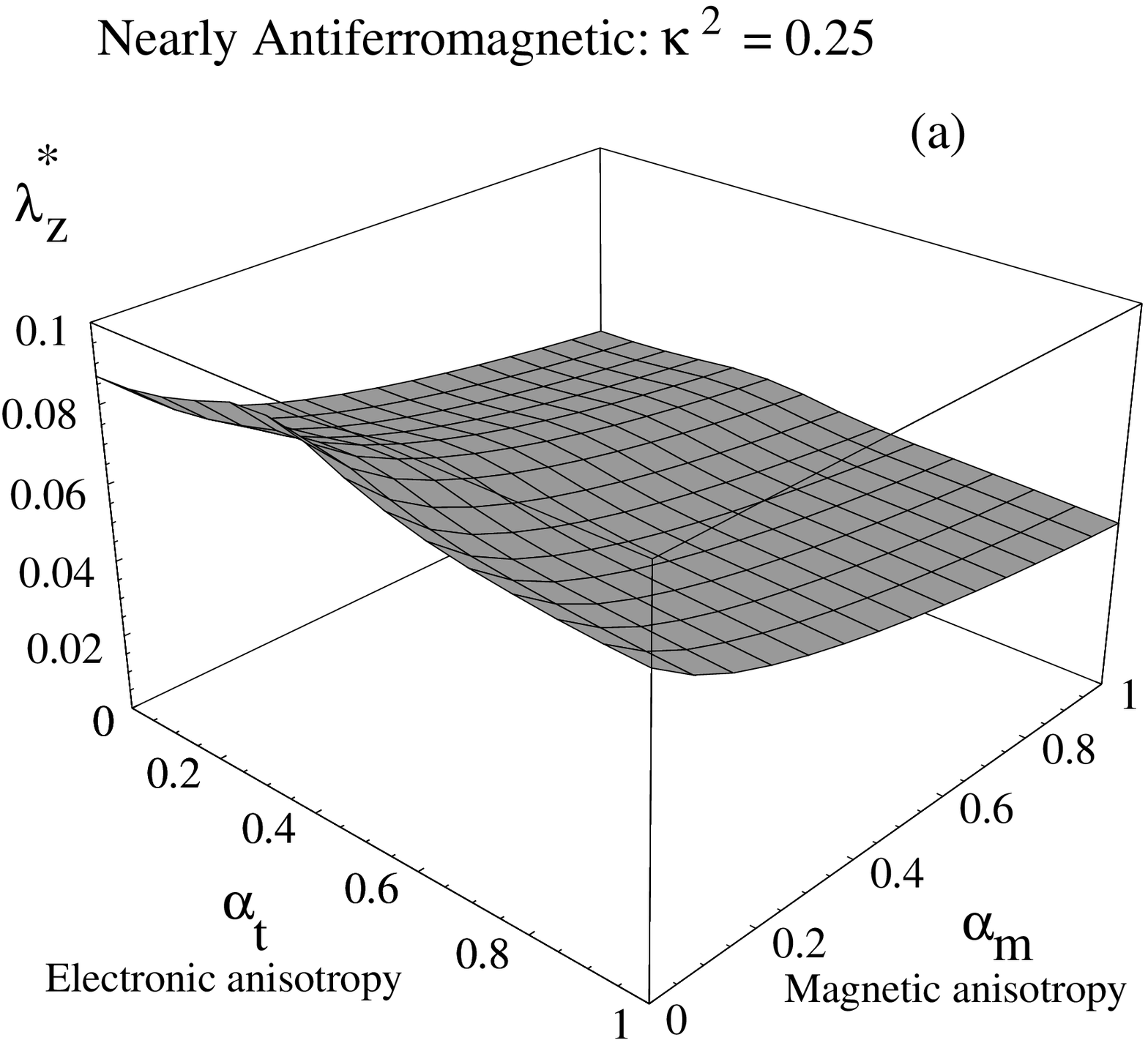}}
\centerline{\epsfysize=6.00in
\epsfbox{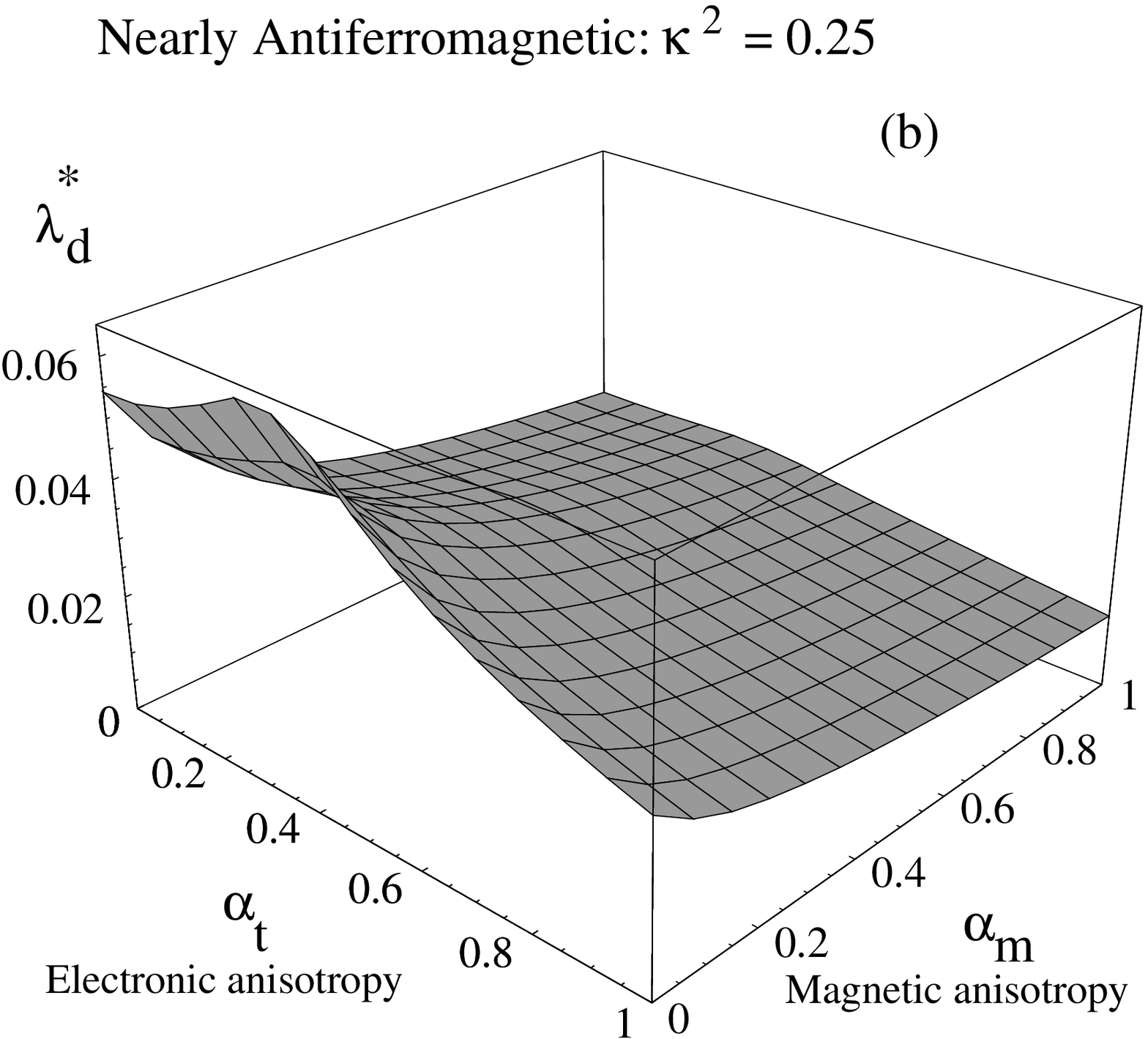}}
\centerline{\epsfysize=6.00in
\epsfbox{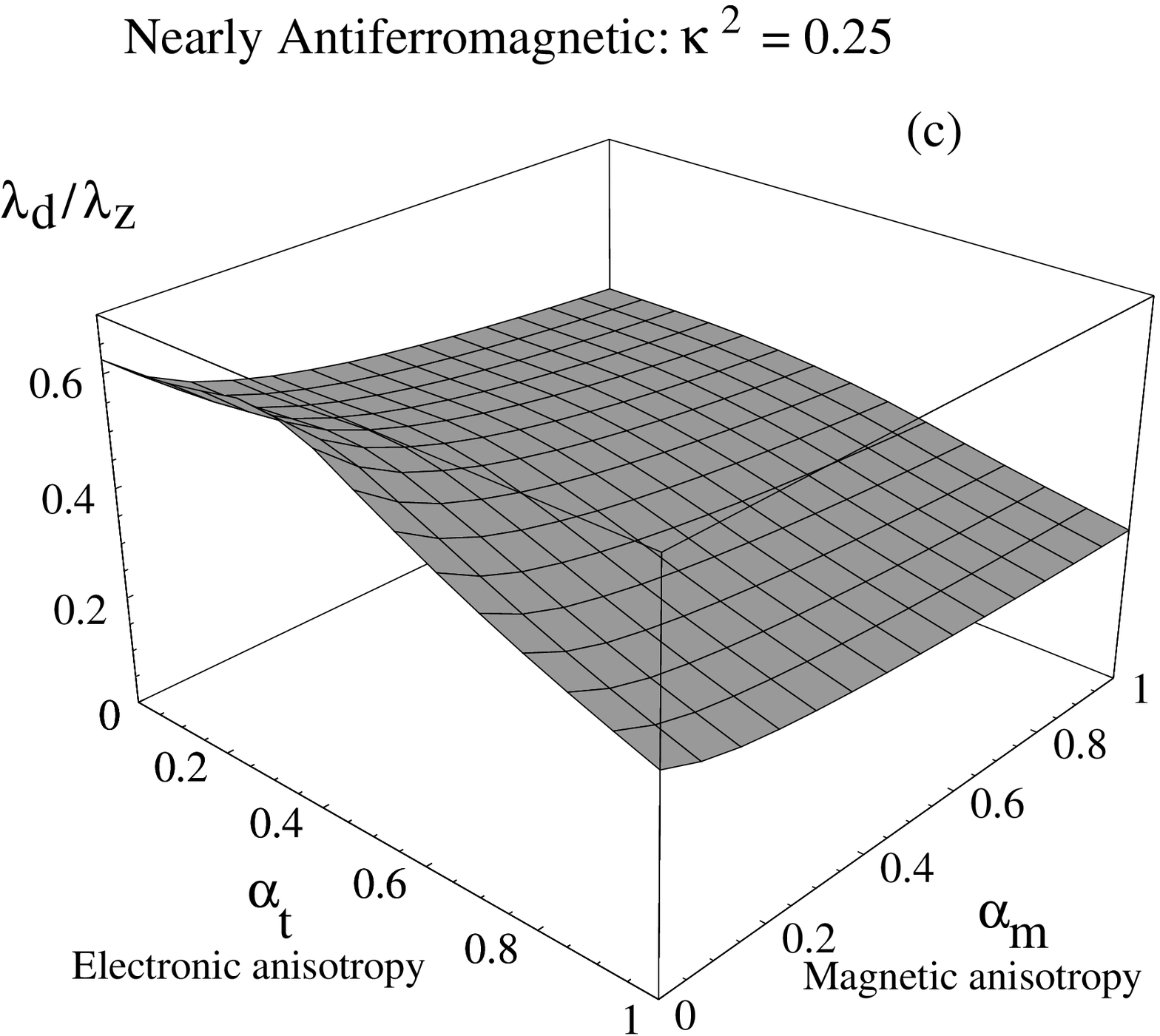}}
\vskip 1.0cm
\caption{Interaction parameters (a) $\lambda_Z^*$, (b) $\lambda_d^*$ and
ratio (c) $\lambda_d/\lambda_Z$ for nearly antiferromagnetic metals for
a representative value of $\kappa^2 = 0.25$}
\label{fig3}
\end{figure}

\begin{figure}
\centerline{\epsfysize=6.00in
\epsfbox{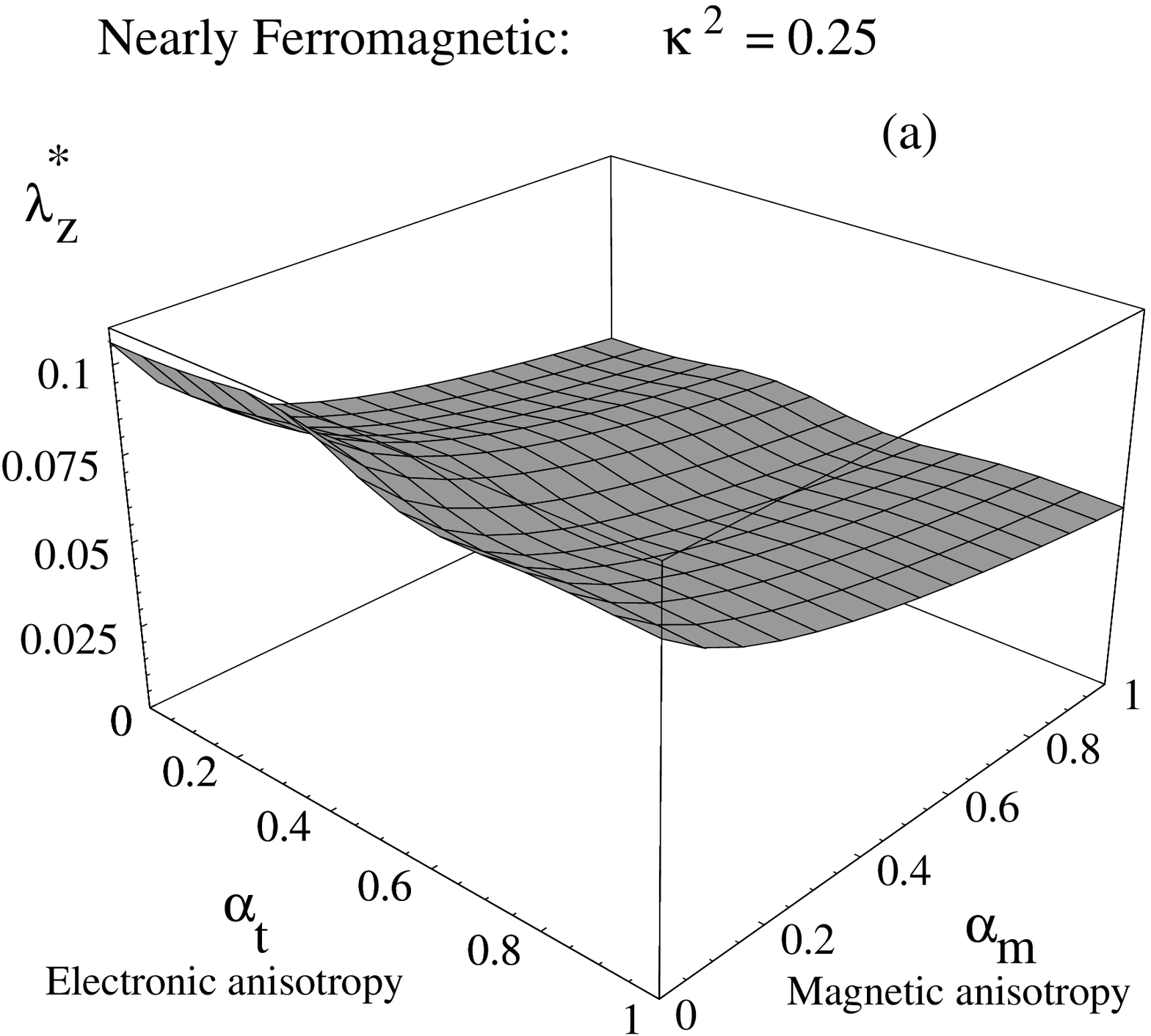}}
\centerline{\epsfysize=6.00in
\epsfbox{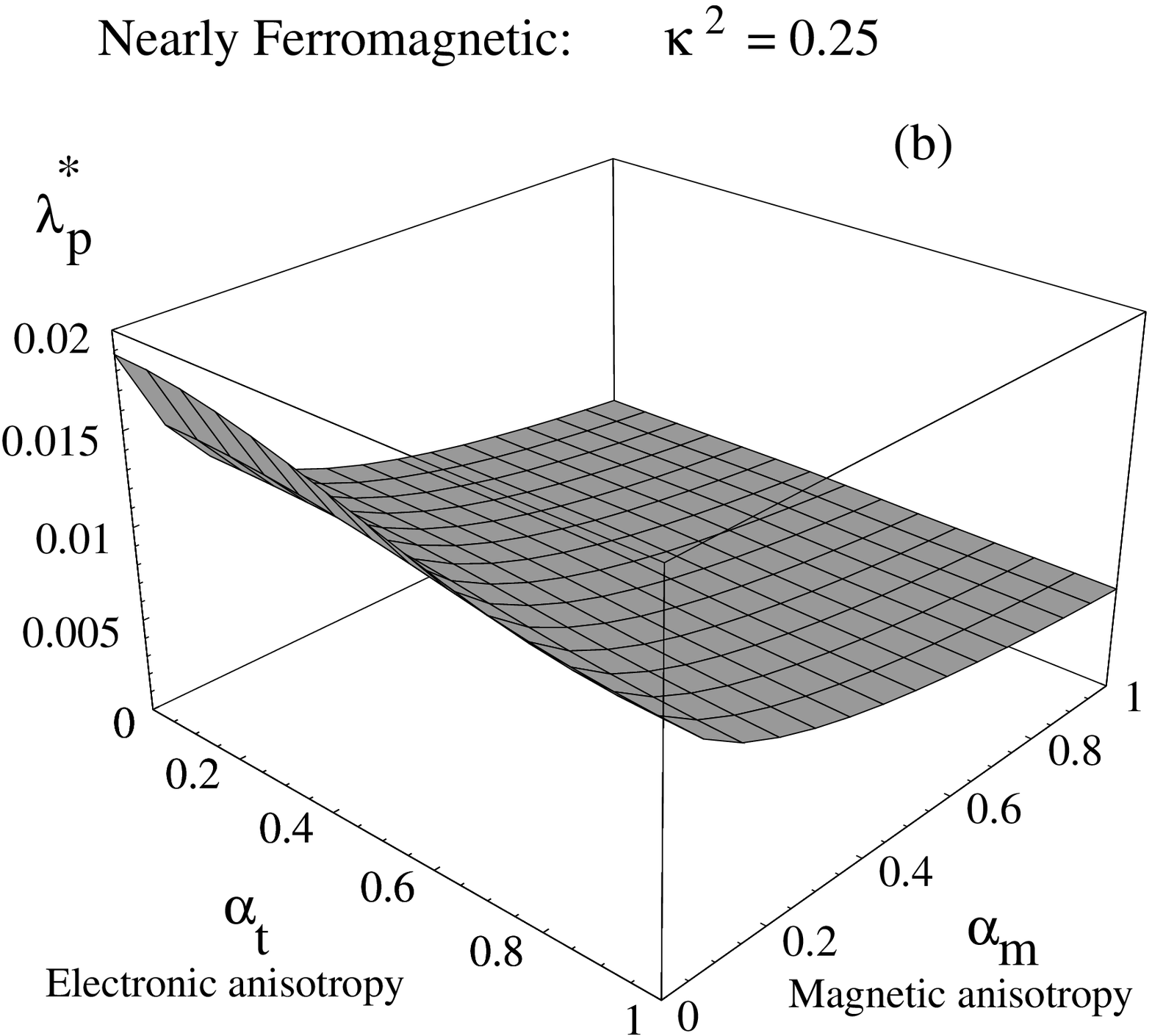}}
\centerline{\epsfysize=6.00in
\epsfbox{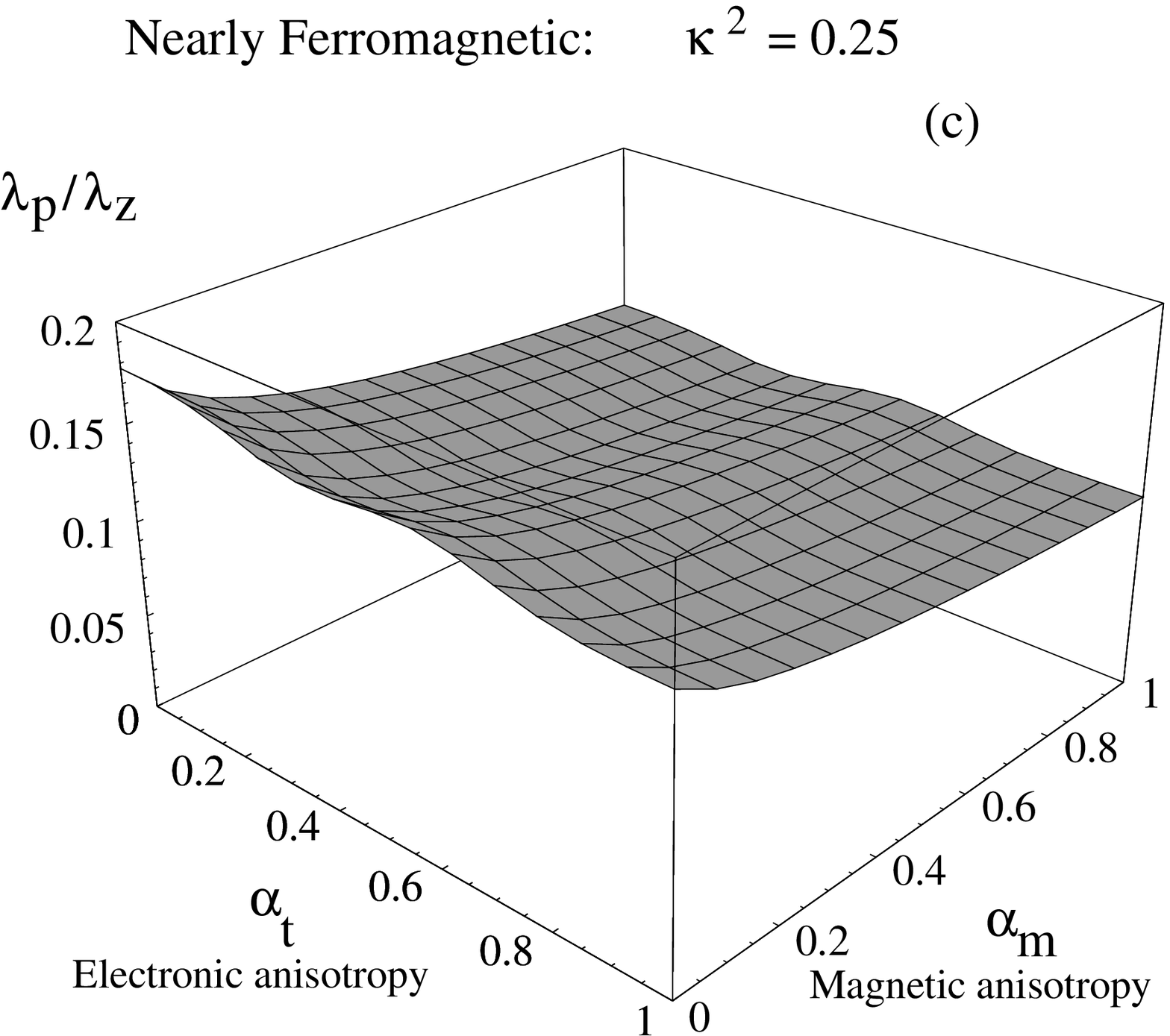}}
\vskip 1.0cm
\caption{Interaction parameters (a) $\lambda_Z^*$, (b) $\lambda_p^*$ and
ratio (c) $\lambda_p/\lambda_Z$ for nearly ferromagnetic metals for
a representative value of $\kappa^2 = 0.25$}
\label{fig4}
\end{figure}

\begin{figure}
\centerline{\epsfysize=6.00in
\epsfbox{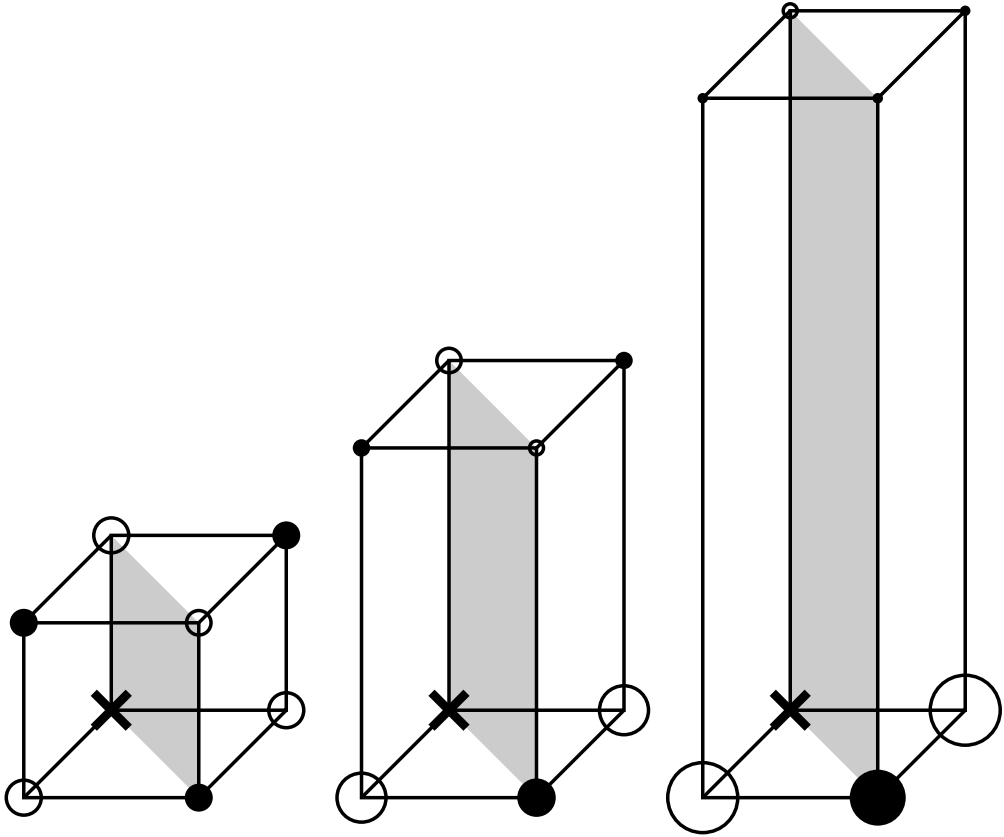}}
\vskip 1.0cm
\caption{The magnetic potential seen by a quasiparticle in a 
spin-singlet $d_{x^2-y^2}$ Cooper pair state given that the other 
quasiparticle is at the origin (marked by a cross).  The figure depicts 
the evolution of the potential as one goes from a cubic to a tetragonal 
lattice by varying the parameter $\alpha_m$.  Closed circles denote 
repulsive sites and open circles attractive ones.  The size of the 
circle is a measure of the strength of the interaction.  
The nodal plane of the $d_{x^2-y^2}$ state are 
represented by the shaded region.}
\label{fig5}
\end{figure}

\begin{figure}
\centerline{\epsfysize=6.00in
\epsfbox{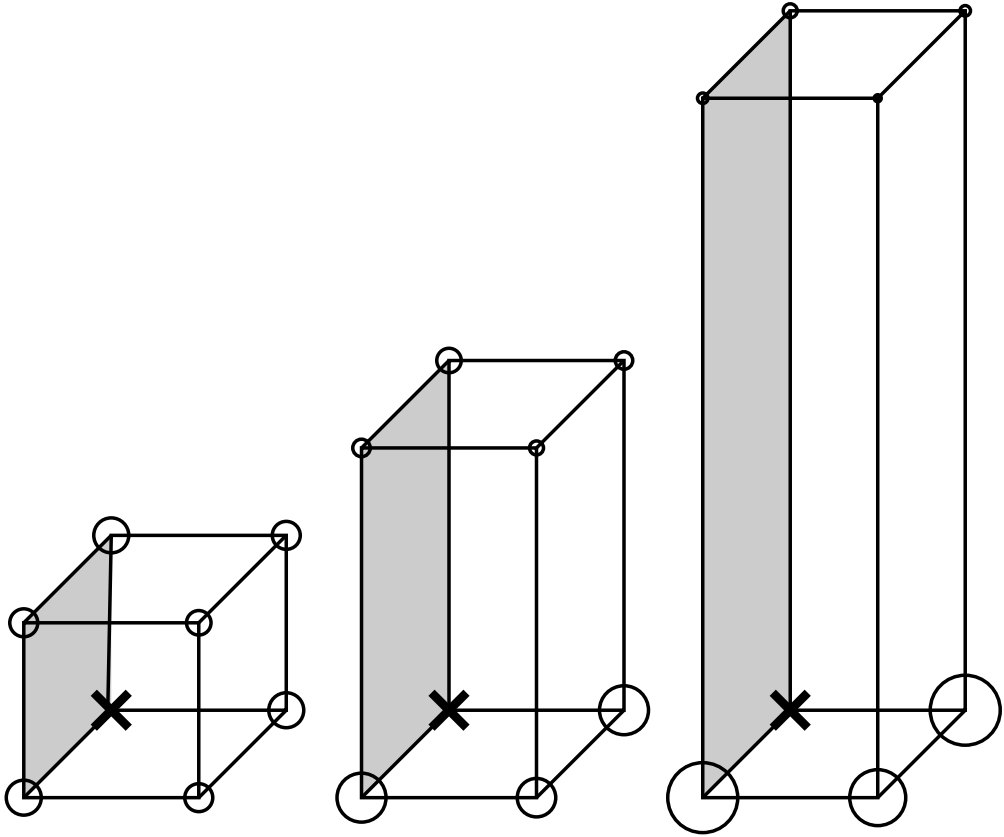}}
\vskip 1.0cm
\caption{The magnetic potential seen by a quasiparticle in a 
spin-triplet $p_{x}$ Cooper pair state given that the other 
quasiparticle is at the origin (marked by a cross).  The figure depicts 
the evolution of the potential as one goes from a cubic to a tetragonal 
lattice by varying the parameter $\alpha_m$.  Open circles denote 
attractive sites. The size of the circle 
is a measure of the strength of the interaction.  The nodal plane of
the $p_{x}$ state is represented by the shaded region.}
\label{fig6}
\end{figure}
\end{document}